\input harvmac.tex
\Title{\vbox{\baselineskip12pt\hbox{hep-th/9711105}
\hbox{RU-97-91}}}
{\vbox{
\centerline{Matrix Description of Heterotic Theory on K3}}}
\vskip 10pt
\centerline{Duiliu-Emanuel Diaconescu and Jaume Gomis}
\medskip
\centerline{\it Department of Physics and Astronomy}
\centerline{\it Rutgers University }
\centerline{\it Piscataway, NJ 08855--0849}
\medskip
\centerline{\tt duiliu, jaume@physics.rutgers.edu}
\medskip
\bigskip
\noindent

The Matrix description of certain $E_8\times E_8$ heterotic theories
on $T^4/Z_2$ is shown to correspond to the world-volume theory of
$Spin(32)/Z_2$ heterotic five-branes on the dual $T^4/Z_2\times S^1$. 
\bigskip

\Date{November 1997}

\newsec{Introduction}

The Discrete Light Cone Quantization description of M-theory as Matrix
theory
\ref\BFSS{T. Banks, W. Fischler, S.H. Shenker, L. Susskind,
Phys.Rev. {\bf D55} (1997) 5112, hep-th/9610043.}
has been derived recently in \ref\SA{N. Seiberg,
hep-th/9710009.}. The techniques applied there for toroidal 
compactifications have been extended to K3 surfaces in 
\ref\DG{D.E. Diaconescu, J. Gomis, hep-th/9710124.}.
The purpose of the present paper is to generalize this derivation 
to K3 compactifications of heterotic string theories. Along the way,
we also provide a derivation of the Matrix description of M-theory
on $T^5/Z_2$ proposed  in 
\nref\Rey{N. Kim, S.-J. Rey, hep-th/9705132.}%
\nref\SB{N. Seiberg,
Phys.Lett. {\bf B408} (1997) 98-104, hep-th/9705221.}%
\refs{\Rey,\SB}. In section two we review the basic principles of \SA\
and outline 
the generalizations to $T^5/Z_2$ and $S^1/Z_2\times T^4/Z_2$.
We explain the occurrence of Gimon-Polchinski models and their 
strong coupling duals. Sections three and four contain a detailed 
analysis applied to nine different heterotic orbifold models.

\newsec{Matrix Theory Derivation}

The essence in deriving the DLCQ description of M-theory is to
consider compactification of M-theory on a light-like circle of radius
$R$ as a
limit of a compactification on a small spatial circle  of radius
$R_s\rightarrow 0$ boosted by a
large amount. The only premise is the assumption that M-theory is
eleven dimensional Lorentz invariant. In order to describe finite
energy states in M-theory one has to rescale the Planck mass of the
theory after boost. In \refs{\SA} this theory was called
$\widetilde{\hbox{M}}$-theory.  A sector with longitudinal momentum
$P^+={N \over R}$ in 
M-theory maps to a sector with momentum ${N \over R_s}$ in
$\widetilde{\hbox{M}}$-theory. Thus, the DLCQ description of M-theory in a
sector with 
$N$ units of longitudinal momentum is described by
$\widetilde{\hbox{M}}$-theory on a small space-like circle with N units of
momentum. This 
corresponds to TypeIIA theory in the regime
where ${\widetilde 
g_s}\rightarrow 
0$ and ${\widetilde M_s} \rightarrow \infty$, whose description is given in
terms of the quantum mechanics of $N$ D0 branes \ref\DKPS{M.R.Douglas,
D. Kabat, P. Pouliot, S.H. Shenker, Nucl.Phys. {\bf B485} (1997)
85, hep-th/9608024.}. This corresponds precisely with the
conjecture made in 
\nref\Suss{L. Susskind, hep-th/9704080.}%
\refs{\BFSS,\Suss}. 

One can consider 
compactifications of M-theory on an arbitrary $d$-dimensional manifold
$X$ in a sector of the theory with $N$ units of longitudinal momentum.
 The longitudinal boost does not affect transverse
coordinates, they just get rescaled with the Planck mass of
$\widetilde{\hbox{M}}$-theory
\eqn\size{
M_P^dV_d=\widetilde{M}_P^d\widetilde{V}_d.}
Thus, the DLCQ description of M-theory on $X$ is given by
$\widetilde{\hbox{M}}$-theory on $S^1\times \widetilde{X}$, where
$S^1$ is a space-like 
circle of vanishing radius $R_s\rightarrow 0$. The size of 
$\widetilde{X}$, 
$\widetilde{V}_d\rightarrow 0$ in the $R_s\rightarrow 0$ limit. One can
conclude that the description is given in terms of $N$ D0 branes
in Type IIA  (as we will see later one can have
 Type IIA orientifolds depending on the choice of
$X$) with vanishing coupling constant and string length on $\widetilde{X}$. 
This theory is better analyzed by performing T-duality $d$-times to
obtain the theory of $N$ D$d$ branes wrapped on the $\it{dual}$ $\widetilde{X}$
manifold of finite size. Following the analysis in \refs{\SA}, one
concludes that the string theory obtained after T-duality is weakly
coupled for $d<3$ and strongly coupled for $d>3$. One can then use
more standard dualities to extract the Matrix theory description of
the compactification.
\subsec{M-theory on $T^5/Z_2$.}
In order to illustrate in a simpler example the method we are
going to use for heterotic Matrix theory, we start by
providing the Matrix description of M-theory on\foot{The low
energy limit of this compactification was first considered in
\ref\FaSm{A. Fayyazuddin, D.J. Smith, Phys.Lett. {\bf B407} (1997) 8,
hep-th/9703208.}.} $T^5/Z_2$. This
example is also interesting since it will provide the Matrix theory
description of Type IIB string theory on $K3$ by the duality in 
\nref\WA{E. Witten, Nucl.Phys. {\bf B463} (1996) 383,
hep-th/9512219.}%
\nref\DM{K. Dasgupta, S. Mukhi, Nucl.Phys. {\bf B465} (1996) 399,
hep-th/9512196.}%
\refs{\WA,\DM}. This example will complete the Matrix theory
description of Type II compactifications on K3.

The DLCQ description of this compactification, following the analysis
in the previous subsection, is that of $\widetilde{\hbox{M}}$-theory on
$S^1\times \widetilde{T}^5/Z_2$, where the radius of $S^1$, $R_s$, shrinks
to zero size. As shown in \refs{\WA} this is dual to a Type IIA
orientifold on $\widetilde{T}^5/Z_2$ with orientifold group
\eqn\orientA{
G=\{1,\Omega R_{56789}\}.}
The $P^+={N \over R}$ sector of M-theory is thus described by $N$
D0-branes on $\widetilde{T}^5/Z_2$ in this Type IIA orientifold. As usual,
one has to perform T-duality on the five circles of the torus,
obtaining Type I theory on\foot{The subscript D
will denote the T-dual manifold.} $\widetilde{T}_D^5$. 
The theory of $N$ D0 branes
is mapped to that of $N$ D5 branes wrapping $\widetilde{T}_D^5$ in Type I
theory. Since we have performed T-duality on more than three
coordinates, the Type I theory is strongly coupled so we must
perform an S-duality transformation. We thus end up with the theory
of $N$  $Spin(32)/Z_2$ heterotic five-branes wrapped on a
five-torus. We have completed the derivation of M-theory on 
$T^5/Z_2$ (and equivalently of Type IIB on K3) and have seen to get
agreement with the proposal in \refs{\SB}. 

Another interesting aspect of this derivation is the identification in
 Matrix theory of the M-theory five-branes that need to be
placed at points in $T^5/Z_2$ to cancel anomalies. By going to the
Type IIA orientifold limit the five-branes become D4-branes localized at
points in $\widetilde{T}^5/Z_2$. After T-duality the D4-brane positions
become Wilson lines in the torus directions, which get mapped to
Wilson lines in the $Spin(32)/Z_2$ heterotic theory after S-duality. 
Thus, Matrix theory encodes the M-theory five-brane positions 
in Wilson lines. 

\subsec{$E_8\times E_8$ Heterotic String on $T^4/Z_2$}

In this subsection we 
outline\foot{We leave the details of the construction for the upcoming 
sections.} the derivation of the
Matrix theory description of the $E_8\times E_8$ 
\nref\Govi{S. Govindarajan, hep-th/9707164.}%
\nref\ReyB{N. Kim, S.-J. Rey,  hep-th/9710192.}%
string\foot{Some six dimensional heterotic compactifications were
considered in \refs{\Govi,\ReyB}.} on
$T^4/Z_2$. The different types of heterotic vacua one can
construct are specified by fixing  the instanton number, $n_1$ and
$n_2$,  
in each $E_8$ gauge
bundle and the number of heterotic five-branes, $n_5$, constrained by the
anomaly cancelation condition $n_1+n_2+n_5=24$ \ref\SW{N. Seiberg,
E. Witten, Nucl.Phys. {\bf B471} (1996) 701,  
hep-th/9603003.}. 
In order to make contact  with the derivation in \refs{\SA}
we will write the heterotic theory as M-theory on
$S^1/Z_2\times T^4/Z_2$ \ref\HW{P. Horava, E. Witten, Nucl.Phys. {\bf B460}
(1996) 506, hep-th/9510209.}. This theory supports an $E_8$ gauge
bundle at the two ends of the interval, with instanton number $n_1$ and
$n_2$, and has $n_5$ M-theory five-branes free to move in the
bulk. The DLCQ description of this theory is given in 
terms of $\widetilde{\hbox{M}}$-theory on $\widetilde{S}^1/Z_2\times
\widetilde{T}^4/Z_2\times S^1$, where the $S^1$ has radius $R_s\rightarrow
0$. The theory obtained by shrinking the circle \refs{\HW} is an orientifold of
Type IIA theory on $\widetilde{S}^1/Z_2\times 
\widetilde{T}^4/Z_2$ with orientifold group
\eqn\orientB{
G=\{1,\Omega R_5, R_{6789}, \Omega R_{56789}\}.}
This theory contains two orientifold
eight-planes at the end of the interval. Each orientifold eight-plane
has 8 superimposed D8 branes which yield an $Spin(16)\times Spin(16)$
gauge group\foot{In \ref\KS{S. Kachru, E. Silverstein,
Phys.Lett. {\bf B396} (1997) 70-76, hep-th/9612162.} it was shown that
in the strong coupling limit (the interval becomes infinite) precisely
the right  degrees of freedom become light to fill out the $E_8\times
E_8$ multiplets.}. This chain of dualities has been considered 
in \ref\CJ{C.V. Johnson, hep-th/9711082.}. Furthermore, within each 
orientifold eight-plane
there are 16 orientifold four-planes, sitting at the singularities in
$\widetilde{T}^4/Z_2$. In order to cancel the D4 brane charge arising from
the orientifold four planes one needs to localize 32 D4 branes at
points in $\widetilde{T}^4/Z_2$. These configurations of orientifold
four-planes and D4 branes inside the D8 branes correspond to the
instantons in the $E_8\times E_8$ picture. The $P^+={N \over R}$
sector is thus described by N D0 
branes on $\widetilde{S}^1/Z_2\times
\widetilde{T}^4/Z_2$ in a Type IIA orientifold with the specified
background. One has to perform T-duality on the five circles which we
 do in two steps. By applying T-duality to the interval
one obtains Type I theory on $\widetilde S^1_D\times \widetilde{T}^4/Z_2$ 
with orientifold group
\eqn\orientC{
G=\{1,\Omega,  R_{6789}, \Omega R_{6789}\},} 
which is  the Gimon-Polchinski (GP) orientifold \ref\GP{E.G. Gimon,
J. Polchinski, Phys.Rev. {\bf D54} (1996) 1667,
hep-th/9601038.}. The D0 branes become D1 strings, the D4 branes
become D5 branes and the D8 branes become D9 branes. This model is
specified by the D5 and D9 brane gauge group\foot{See section four 
for details.} $G_5\times G_9$.  We still have to T-dual
along the four sides of 
the torus since they are very small. By applying T-duality $T_{6789}$
we get back the GP model with orientifold group \orientC. This T-duality
exchanges $D5\leftrightarrow D9$ and the D1 string becomes a D5 brane
of Type I wrapped on the compact directions. The two types of D5
branes intersect along a
circle. Since we have performed T-duality on five circles, the Type I 
theory we have is strongly coupled, so we must S-dual
 to get to heterotic $Spin(32)/Z_2$. We conclude that
the $P^+={N \over R}$ sector of the $E_8\times E_8$ heterotic string 
compactification  we are considering has as a Matrix description the
theory of $N$ $Spin(32)/Z_2$ five-branes wrapped on $\widetilde{S}^1_D\times
\widetilde{T}^4_D/Z_2$. 

The Matrix theory description of the $E_8\times E_8$ compactification
also encodes the background Wilson lines of the heterotic theory. They
can be traced by  the chain of dualities we have
followed. The Wilson lines in the $E_8\times E_8$ theory can be pushed
to the M-theory limit (see footnote 4). 
In the Type
IIA orientifold limit, $E_8\times E_8$ Wilson
lines remain Type
IIA Wilson lines. T-duality on
the first circle also leaves them unaffected. The final
$T_{6789}$ transformation maps the Wilson lines to D5-brane positions
which after S-duality are heterotic five-brane positions. It is
important to note that these five-branes are the ones localized at
points on
$\widetilde{T}^4_D/Z_2$  and wrapping $\widetilde{S}^1_D$. 
Thus,
$E_8\times E_8$ Wilson lines get mapped to five-brane positions in
$\widetilde{T}^4_D/Z_2$ in the
Matrix theory description. 

One can try to give a Matrix description of nonperturbative
$E_8\times E_8$ vacua by putting $n_5$ M-theory five-branes in the
bulk. In the Type IIA  orientifold limit these become D4-branes
localized in $\widetilde{S}^1\times\widetilde{T}^4/Z_2$. After T-duality on
the interval they correspond to D5-branes wrapping the circle and
localized in  $\widetilde{T}^4/Z_2$. Performing the remaining part of the
T-duality maps the positions of the branes to Wilson lines on
$\widetilde{T}^4_D/Z_2$. The final S-duality transformation preserves these
Wilson lines. Thus $E_8\times E_8$ five-branes get mapped to Wilson
lines along $\widetilde{T}^4_D/Z_2$ in Matrix theory. 
Matrix theory  captures in an interesting way both the
Wilson lines and  five-brane positions of the $E_8\times E_8$
theory.

\newsec{The space-time theories}

Here we present a detailed construction of the $E_8\times E_8$
heterotic theories mentioned in
the previous section. All theories can be understood 
starting from a basic nonperturbative $T^4/Z_2$ 
orbifold construction 
\nref\ibanezA{G. Aldazabal, A. Font, L.E. Iba\~nez, A.M. Uranga,
G. Violero, hep-th/9706158.}%
\refs{\ibanezA}. Since we will be ultimately concerned with orbifolds with 
Wilson lines, we briefly review the relevant consistency conditions.
The embedding of the space group $Z_2$ in the $E_8\times E_8$ gauge group 
is defined by a shift vector $V^I$, where $I=1,\ldots,16$ is a
$\Gamma_8\times \Gamma_8$ lattice index.
The Wilson lines can be regarded similarly as shift vectors
$A_i^I, i=6,\ldots,9$. They must satisfy constraints 
\nref\dhvw{L. Dixon, J.A. Harvey, C. Vafa, E. Witten, Nucl.Phys.
{\bf B261} (1985) 678; {\bf B274} (1986) 285.}%
\nref\ibanezB{L.E. Iba\~nez, H.-P. Nilles, F. Quevedo,
Phys.Lett. {\bf B192} (1987) 332.}%
\nref\ibanezC{L.E. Iba\~nez, J. Mas, H.-P. Nilles, F. Quevedo,
Nucl.Phys. {\bf B301} (1988) 157.}%
\refs{\dhvw, \ibanezB, \ibanezC}
imposed by the group action
\eqn\constA{
2V^I\in\Gamma_8\times \Gamma_8,}
\eqn\constB{
2A_i^I\in\Gamma_8\times \Gamma_8,}
\eqn\constC{
2\left(\sum_{I=1}^{16}(A_i^I)^2\right)\equiv 0\ (\hbox{mod 2}),}
\eqn\constD{
2\left(\sum_{I=1}^{16}A_i^IA_j^I\right)\equiv 0\ (\hbox{mod 1}),\qquad
i\neq j,}
\eqn\constD{
2\left(\sum_{I=1}^{16}V^IA_i^I\right)\equiv 0\ (\hbox{mod 1}), 
\qquad i=6,\ldots,9,}
and modular invariance
\eqn\constE{
2\left(\sum_{I=1}^{16}(nV^I+r_iA_i^I)^2-n^2v^2\right)\equiv 0\ 
(\hbox{mod 2}),}
where $n=0,1,\ r_i=0,1$ and $v=({1\over 2}, -{1\over 2})$ is the
vector of $Z_2$ charges of the toroidal coordinates.

Nonperturbative orbifolds
\ibanezA\ are heterotic vacua described by a perturbative part 
consisting of a free orbifold and a nonperturbative part, in general
background five-branes. The free orbifold does not satisfy the modular
invariance conditions, therefore it has an anomalous perturbative spectrum.
The anomaly is then cancelled by taking into account the contribution
of the five-branes as in 
\refs{\SW}.
The theories of interest here can be constructed starting from 
a nonperturbative heterotic orbifold which is on the same moduli 
as the $(8,8)$ smooth compactification with $8$ five-branes. 
The shift vector is given by
\eqn\shiftA{V={1\over 4}(1,1,\ldots 1)\times{1\over 4}
(1,1,\ldots,1).}
It is easy to see that this violates modular invariance. The 
anomalous spectrum contains an $E_7\times SU(2)\times E_7\times SU(2)$
gauge group with hypermultiplets in 
$2({\bf 56},{\bf 2};{\bf 1},{\bf 1})\oplus 2({\bf 1},{\bf 1};
{\bf 56},{\bf 2})\oplus 4({\bf 1},{\bf 1};{\bf 1},{\bf 1})$
from the untwisted sector and 
$8({\bf 1},{\bf 2};{\bf 1},{\bf 1})\oplus
8({\bf 1},{\bf 1};{\bf 1},{\bf 2})\oplus 
16({\bf 1},{\bf 1};{\bf 1},{\bf 1})$
from the twisted sector. 
It can be checked that adding eight five-branes renders the
spectrum anomaly free.
Higgsing the $SU(2)$ factors by giving vacuum expectation values
to the doublet
hypermultiplets yields the spectrum of the corresponding smooth 
compactification. 

Before giving an M-theory description of this model we consider the
problem of Wilson lines. It is clear from \constA-\constD\ that these
must be quantized. There are three cases of interest 
which are presented in Tables A1-A3.

In finite coupling regime the higgsed model is described as M-theory 
on $S^1/Z_2\times T^4/Z_2$ 
\refs{\HW}.
The two $E_8$ vector bundles on the nine dimensional planes have
instanton number $8$ on $T^4/Z_2$. The instantons are small singular 
configurations of charge ${1\over 2}$ hidden at the sixteen fixed
points. As explained in 
\nref\BLP{M. Berkooz, R.G. Leigh, J. Polchinski, J.H. Schwarz,
N. Seiberg, E. Witten,  Nucl.Phys. {\bf B475} (1996) 115.}%
\refs{\BLP, \ibanezA},
instantons hidden at $C^2/Z_2$ singularities are classified by all
possible embeddings of $Z_2$ in a $Spin(16)/Z_2$ subgroup of $E_8$. 
The case of relevance here is that of instantons without vector
structure. There are two possible embeddings differing in the action 
on spinors called $N$ and $N^\prime$ in \BLP. In lattice language
these correspond to shift vectors $Q={1\over 4}(1,\ldots,1)$ 
and $Q^\prime={1\over 4}(1,\ldots,-3)$. The instanton charge is 
respectively $k\equiv {1\over 2}\ (\hbox{mod Z})$ and $k\equiv 
{1\over 4}\  \hbox{(mod Z)}$ in the two cases\foot{The values are 
$1$ and $1/2$ for $Spin(32)/Z_2$ instantons
but they are halved for $Spin(16)/Z_2$ instantons. This can be easily
seen from the fact that we have to embed only eight copies of the 
basic $U(1)$ instanton field \BLP\ in the gauge group rather 
than sixteen.}.
Therefore the first case describes instantons hidden at bare orbifold
singularities (no fixed five-branes). As detailed below there are more
general configurations related to the second embedding.

Note that a 
mobile five-brane really is a unit of four five-branes identified 
by the two $Z_2$ projections. Therefore they can also be distributed in
configurations with fixed five-branes at the singularities. 
A single five-brane stuck at the singularity contributes an instanton 
number $1\over 4$, thus corresponding to the second case above. 
The possible configurations are restricted
by nonperturbative anomalies of the type considered in \BLP.
Since these are of topological nature we can use their arguments
directly. There are three allowed configurations which combined with
the three sets of Wilson lines give nine different models presented in
Tables A1-A3.
Note that the second series of models is on the same moduli as the smooth 
$(12,12)$ compactification and it has been considered in a very
similar context in \refs{\CJ}. The third
series of models corresponds to the smooth $(10,10)$ case with four 
five-branes in the bulk. The two groups of eight quarter five-branes 
must lie in identical three planes on the two walls.

\newsec{The auxiliary theories}

In this section we present the Matrix description of the heterotic
$E_8\times E_8$ orbifolds constructed above. The extension of
Seiberg's derivation \SA\ to the present situation has been discussed
in section two. The relevant chain of dualities leads to a
strong-coupling limit of the Gimon-Polchinski orientifolds. 
To understand the precise relation between the initial
light cone heterotic 
theories and the orientifolds, we will describe 
in some detail a particular example, namely the model 2.1 in Table A2. 
There are no Wilson lines on $T^4/Z_2$, but we must turn on a 
Wilson line on the shrinking circle in order to connect with the 
Type IIA theories which have unbroken gauge group 
$Spin(16)\times Spin(16)$. The $Spin(16)$ gauge bundle on the
D8-branes at one end of the interval has sixteen small instantons
of charge $3/4$ hidden at the singularity. These can be associated
with the four-dimensional orientifold planes ($1/2$) and to quarter
D4-branes fixed at the singularities ($1/4$). The gauge group is
broken to $U(8)\times U(8)$. 
T-duality on the interval maps the Type IIA theory to a GP model with 
sixteen half five-branes at the fixed points. Therefore two $Spin(16)/Z_2$ 
instantons located at identical fixed points on the two walls are 
merged into 
a single $Spin(32)/Z_2$ instanton without vector structure and 
with double the charge (see footnote 5).
The dual of the 
interval is an extra compact circle with a Wilson line breaking the
D9-brane gauge group\foot{This is the classical gauge group. 
At quantum level, sixteen $U(1)$ factors of the total gauge group 
are broken by anomalies \BLP.}
$U(16)$ to $U(8)\times U(8)$. T-duality on the
remaining four compact coordinates yields another GP model with 
Wilson lines on $T^4/Z_2$ breaking the gauge group to $U(1)^{16}$.
These are dual to the half D5-branes of the original model. On the
other hand, the Wilson lines of the original model are mapped to 
D5-brane positions in the new theory. In the case studied here 
this means that we are finally left with eight D5-branes collapsed
at one singularity giving an $U(16)$ gauge group. This is 
further broken by the Wilson line on the circle to $U(8)\times U(8)$. 
The other cases can be treated similarly. The $E_8\times E_8$ Wilson
lines in Tables A1-A3 give $Spin(32)/Z_2$ Wilson lines
in the GP model by merging together the eight dimensional vectors. 
Then, as a general rule, the 
five-brane configuration of the final GP model is determined by 
the Wilson lines of the original model and vice-versa.
Note that the GP Wilson lines derived this way are precisely the 
three sets of Wilson lines satisfying the consistency conditions 
of GP models \BLP. Therefore the above chain of dualities connects
the nine heterotic theories with the nine consistent
GP theories found in \BLP. 

As explained in  section two, the final step in the Matrix model 
derivation is to take the strong coupling limit of the GP theories. 
This is done by constructing the S-dual weakly coupled heterotic 
$Spin(32)/Z_2$ theories. Therefore, we expect nine different heterotic 
auxiliary models corresponding to the nine $E_8\times E_8$ theories 
that we started with. 
The easiest case is when there are sixteen half five-branes at the 
singularities. Then the $Spin(32)/Z_2$ models are free orbifold theories
without vector structure
\nref\ibanezD{G. Aldazabal, A. Font, L.E. Iba\~nez, F. Quevedo,
Phys.Lett. {\bf B380} (1996) 33, hep-th/9602097.}%
\refs{\ibanezD, \BLP}.

The remaining situations are

\item{\it i)}{\it eight five-branes in the bulk}. In this case the
heterotic dual models are nonperturbative orbifolds constructed in 
\ibanezA. As expected there are eight heterotic $Spin(32)$ five-branes
present in order to cancel the anomalies.

\item{\it ii)}{\it eight half-five branes at eight singularities in a
three plane and four mobile five-branes in the bulk}. This case is not
known. We propose that it can be obtained from the previous one by 
splitting four of the bulk heterotic five-branes in eight halves
localized at eight singularities in the corresponding three plane.

\noindent
The results are summarized in Tables B1-B3. Note that the models
are ordered so that they correspond precisely to the $E_8\times E_8$ 
theories in Tables A1-A3. More precisely, this means that the 
Matrix model for a heterotic theory in Tables A1-A3 is formulated in terms
of $N$ heterotic-five branes embedded in the corresponding auxiliary
heterotic theory in Tables B1-B3.
Some remarks are in order here:
\item{\it i)} It turns out that the Matrix description of $E_8\times E_8$
heterotic theories is formulated in terms of $Spin(32)$ five-branes. 
The relation between the space-time and base theory gauge bundles 
is determined essentially by T-duality as described in the adjoint
tables. 

\item{\it ii)} The auxiliary heterotic theories in which  the
base theories are embedded have in many cases background five-branes
filling the 
noncompact directions. They intersect the wrapped five-branes 
along strings wound around $S^1$. These strings should 
be regarded\foot{We thank Nathan Seiberg for explaining this to us.}
as impurities or sources in the world-volume theories of the
five-branes\foot{Similar analysis of impurities in little Matrix
theory was done in \ref\Sethi{S. Sethi, hep-th/9710005.}.}. The degrees
of freedom on these background strings  
can be easily determined in the GP picture. One obtains chiral 
fermions transforming in the bifundamental of the dynamical gauge
group on the wrapped five-branes and the space-time gauge
group determined by transverse five-branes. 
This mechanism explains the global symmetry of the base theory
associated with background small instantons.

\item{\it iii)} In all the cases listed below, whenever mobile five-branes
are present, they are assumed collapsed at a single orbifold
singularity. Such configurations are T-dual to trivial Wilson lines. 
If the five-branes move away from the fixed points in the original 
$E_8\times E_8$ models, this should correspond to turning on continuous
Wilson lines of the form
\eqn\contwls{
W=\hbox{diag}(R(\theta_i))}
in the auxiliary theories,
where $R(\theta_i)$ are $2\times 2$ rotation matrices \BLP.
This cannot be achieved within the framework of free 
$Spin(32)/Z_2$ orbifold theories
that we considered. The deformations should be regarded as 
exactly marginal operators of the free theories 
which move the model away from the free point in the CFT moduli space. 

\newsec{Discussion}

In this paper we have derived the Matrix description of M-theory on
$T^5/Z_2$ and $S^1/Z_2\times T^4/Z_2$. The derivation  maps in an 
explicit manner the
parameters of  the space-time theories with those of the auxiliary
theories. The appearance of heterotic five-branes is manifest in our
constructions. 

The Matrix description of M-theory on $T^5/Z_2$ also provides a
description of Type IIB on K3 \refs{\WA,\DM}. This is given in terms of
heterotic five-branes wrapped on $T^5$, a theory with 8 real
supercharges. This example completes the Matrix  description of
Type II theories on K3.

We presented Matrix descriptions of certain $E_8\times E_8$
heterotic vacua in the orbifold limit of K3. These are given in terms
of heterotic five-branes wrapped on $\widetilde{S}^1_D\times
\widetilde{T}^4_D/Z_2$, which are theories with 4 real
supercharges. The reduced number of supersymmetries has not allowed us
to give a precise description of BPS states as in \refs{\SB,\DG}. 
This is correlated with the known difficulties in the study of
BPS spectra of conventional string theories with eight space-time 
supercharges.
The methods we use
can even provide us with Matrix descriptions of nonperturbative
heterotic vacua. Many more cases can be studied. One can consider
vacua with instantons embedded in an asymmetrical way between  the gauge
bundles. A very challenging problem would be to provide Matrix
descriptions of arbitrary smooth K3 compactifications.
\vfill\eject

\bigskip
\centerline{\bf Acknowledgments}
We are very grateful to Nathan Seiberg for suggesting the problem and
for collaboration at various stages of the project. We would also like
to thank Ofer Aharony, Tom Banks, Micha Berkooz, Rami Entin and Steve 
Shenker for helpful discussions.

\bigskip\noindent
{\it Note added}

\noindent
In course of completion of this work the paper \CJ\ with similar
subject and conclusions appeared.

\vfill\eject

\midinsert{\vbox{\qquad \qquad \qquad \qquad \qquad
Table A1. $E_8\times E_8$ heterotic orbifolds. 
$$\vbox{
\offinterlineskip
\halign{
\strut \vrule \quad $#$\quad &\vrule \hfill \quad #\quad \hfill
&\vrule \quad $#$\quad &\vrule \hfil \quad #\quad \hfill \vrule \cr
\noalign{\hrule}
\hbox{Model} & Shift vector & \hbox{Wilson lines} & Five-branes\cr
\noalign{\hrule}
1.1 & $\eqalign{&\qquad\cr&V={1\over 4}(1\ldots 1)\times\cr
&\qquad\ {1\over 4}(1\ldots 1)\cr &\qquad\cr}$ &
\qquad\qquad\hbox{none} & $\eqalign{& \hbox{8 mobile}\cr & 
\quad \hbox{5B}\cr}$ \cr
\noalign{\hrule}
1.2 & $\eqalign{&\qquad\cr&V={1\over 4}(1\ldots 1)\times\cr
&\qquad\ {1\over 4}(1\ldots 1)\cr &\qquad\cr}$ &
\eqalign{&\qquad\cr& A_6={1\over 2}(1,1,1,0,0,0,1,0)\times\cr
&\qquad\ {1\over 2}(1,1,1,0,0,0,1,0)\cr 
& A_7={1\over 2}(1,1,0,1,0,1,0,0)\times\cr
&\qquad\ {1\over 2}(1,1,0,1,0,1,0,0)\cr
& A_8={1\over 2}(1,0,1,1,1,0,0,0)\times\cr
&\qquad\ {1\over 2}(1,0,1,1,1,0,0,0)\cr
& A_9={1\over 2}(1,1,1,1,1,1,1,1)\times\cr
&\qquad\ {1\over 2}(0,0,0,0,0,0,0,0)\cr &\qquad\cr
} & $\eqalign{& \hbox{8 mobile}\cr & \quad \hbox{5B}\cr}$ \cr
\noalign{\hrule}
1.3& $\eqalign{&V={1\over 4}(1\ldots 1)\times\cr
&\qquad\ {1\over 4}(1\ldots 1)\cr}$ &
\eqalign{&\ \cr& A_6={1\over 2}(1,1,0,0,0,0,0,0)\times\cr
&\qquad\ {1\over 2}(1,1,0,0,0,0,0,0)\cr 
& A_7={1\over 2}(1,0,1,0,0,0,0,0)\times\cr
&\qquad\ {1\over 2}(1,0,1,0,0,0,0,0)\cr
& A_8={1\over 2}(1,1,0,0,0,0,0,0)\times\cr
&\qquad\ {1\over 2}(1,1,0,0,0,0,0,0)\cr
& A_9={1\over 2}(0,0,0,0,0,0,0,0)\times\cr
&\qquad\ {1\over 2}(0,0,0,0,0,0,0,0)\cr &\ \cr}
&
$\eqalign{& \hbox{8 mobile}\cr & \quad \hbox{5B}
\cr}$ \cr
\noalign{\hrule}
}
}$$
}}
\endinsert
\vfill\eject
\midinsert{\vbox{\qquad \qquad \qquad \qquad \qquad
Table A2. $E_8\times E_8$ heterotic orbifolds.
$$\vbox{
\offinterlineskip
\halign{
\strut \vrule \quad $#$\quad &\vrule \hfill \quad #\quad \hfill
&\vrule \quad $#$\quad &\vrule \hfil \quad #\quad \hfill \vrule \cr
\noalign{\hrule}
\hbox{Model} & Shift vector & \hbox{Wilson lines} & Five-branes \cr
\noalign{\hrule}
2.1 & $\eqalign{&\ \cr&V={1\over 4}(1\ldots 1)\times\cr
&\qquad\ {1\over 4}(1\ldots 1)\cr &\ \cr}$ &
\qquad\qquad\hbox{none} & $\eqalign{& 16\times {1\over 2}\ 
\hbox{5B}\cr & 
\hbox{on each wall}\cr}$ \cr
\noalign{\hrule}
2.2 & $\eqalign{&V={1\over 4}(1\ldots 1)\times\cr
&\qquad\ {1\over 4}(1\ldots 1)\cr}$ &
\eqalign{&\ \cr& A_6={1\over 2}(1,1,1,0,0,0,1,0)\times\cr
&\qquad\ {1\over 2}(1,1,1,0,0,0,1,0)\cr 
& A_7={1\over 2}(1,1,0,1,0,1,0,0)\times\cr
&\qquad\ {1\over 2}(1,1,0,1,0,1,0,0)\cr
& A_8={1\over 2}(1,0,1,1,1,0,0,0)\times\cr
&\qquad\ {1\over 2}(1,0,1,1,1,0,0,0)\cr
& A_9={1\over 2}(1,1,1,1,1,1,1,1)\times\cr
&\qquad\ {1\over 2}(0,0,0,0,0,0,0,0)\cr
&\ \cr} & $\eqalign{& 16\times {1\over 2}\ \hbox{5B}\cr & 
\hbox{on each wall}\cr}$ \cr
\noalign{\hrule}
2.3& $\eqalign{& V={1\over 4}(1\ldots 1)\times\cr
&\qquad\ {1\over 4}(1\ldots 1)\cr}$ &
\eqalign{&\ \cr& A_6={1\over 2}(1,1,0,0,0,0,0,0)\times\cr
&\qquad\ {1\over 2}(1,1,0,0,0,0,0,0)\cr 
& A_7={1\over 2}(1,0,1,0,0,0,0,0)\times\cr
&\qquad\ {1\over 2}(1,0,1,0,0,0,0,0)\cr
& A_8={1\over 2}(1,1,0,0,0,0,0,0)\times\cr
&\qquad\ {1\over 2}(1,1,0,0,0,0,0,0)\cr
& A_9={1\over 2}(0,0,0,0,0,0,0,0)\times\cr
&\qquad\ {1\over 2}(0,0,0,0,0,0,0,0)\cr &\ \cr}
&
$\eqalign{& 16\times {1\over 2}\ \hbox{5B}\cr & 
\hbox{on each wall}
\cr}$ \cr
\noalign{\hrule}
}
}$$
}}
\endinsert
\vfill\eject
\midinsert{\vbox{\qquad \qquad \qquad \qquad \qquad
Table A3. $E_8\times E_8$ heterotic orbifolds.
$$\vbox{
\offinterlineskip
\halign{
\strut \vrule \quad $#$\quad &\vrule \hfill \quad #\quad \hfill
&\vrule \quad $#$\quad &\vrule \hfil \quad #\quad \hfill \vrule \cr
\noalign{\hrule}
\hbox{Model} & Shift vector & \hbox{Wilson lines} & Five-branes \cr
\noalign{\hrule}
3.1 & $\eqalign{&\ \cr&V={1\over 4}(1\ldots 1)\times\cr
&\qquad\ {1\over 4}(1\ldots 1)\cr &\ \cr}$ &
\qquad\qquad\hbox{none} & $\eqalign{& 8\times {1\over 4}\ 
\hbox{5B}\cr  
&\hbox{in a 3-plane}\cr
&\hbox{on each wall,}\cr &\hbox{4 mobile}\ \hbox{5B}\cr}$ \cr
\noalign{\hrule}
3.2 & $\eqalign{&V={1\over 4}(1\ldots 1)\times\cr
&\qquad\ {1\over 4}(1\ldots 1)\cr}$ &
\eqalign{&\ \cr& A_6={1\over 2}(1,1,1,0,0,0,1,0)\times\cr
&\qquad\ {1\over 2}(1,1,1,0,0,0,1,0)\cr 
& A_7={1\over 2}(1,1,0,1,0,1,0,0)\times\cr
&\qquad\ {1\over 2}(1,1,0,1,0,1,0,0)\cr
& A_8={1\over 2}(1,0,1,1,1,0,0,0)\times\cr
&\qquad\ {1\over 2}(1,0,1,1,1,0,0,0)\cr
& A_9={1\over 2}(1,1,1,1,1,1,1,1)\times\cr
&\qquad\ {1\over 2}(0,0,0,0,0,0,0,0)\cr
&\ \cr} & $\eqalign{& 8\times {1\over 4}\ \hbox{5B}\cr  
&\hbox{in a 3-plane}\cr
&\hbox{on each wall,}\cr &\hbox{4 mobile}\ \hbox{5B}\cr}$ \cr
\noalign{\hrule}
3.3& $\eqalign{&V={1\over 4}(1\ldots 1)\times\cr
&\qquad\ {1\over 4}(1\ldots 1)\cr}$ &
\eqalign{&\ \cr& A_6={1\over 2}(1,1,0,0,0,0,0,0)\times\cr
&\qquad\ {1\over 2}(1,1,0,0,0,0,0,0)\cr 
& A_7={1\over 2}(1,0,1,0,0,0,0,0)\times\cr
&\qquad\ {1\over 2}(1,0,1,0,0,0,0,0)\cr
& A_8={1\over 2}(1,1,0,0,0,0,0,0)\times\cr
&\qquad\ {1\over 2}(1,1,0,0,0,0,0,0)\cr
& A_9={1\over 2}(0,0,0,0,0,0,0,0)\times\cr
&\qquad\ {1\over 2}(0,0,0,0,0,0,0,0)\cr &\ \cr}
&
$\eqalign{& 8\times {1\over 4}\ \hbox{5B}\cr
&\hbox{in a 3-plane}\cr  
&\hbox{on each wall,}\cr & \hbox{4 mobile}
\ \hbox{5B}\cr}$ \cr
\noalign{\hrule}
}
}$$
}}
\endinsert
\vfill\eject
\midinsert{\vbox{\qquad \qquad \qquad \qquad \qquad
Table B1. $Spin(32)/Z_2$ heterotic orbifolds.
$$\vbox{
\offinterlineskip
\halign{
\strut \vrule \quad $#$\quad &\vrule \hfill \quad #\quad \hfill
&\vrule \quad $#$\quad &\vrule \hfil \quad #\quad \hfill \vrule \cr
\noalign{\hrule}
\hbox{Model} & Shift vector & \hbox{Wilson lines} & Five-branes \cr
\noalign{\hrule}
1.1 & $V={1\over 4}(1\ldots 1)$ &
\quad\hbox{none} & $\eqalign{& \hbox{8 mobile}\cr & 
\quad \hbox{5B}\cr}$ \cr
\noalign{\hrule}
1.2 & $\eqalign{&\ \cr
& V={1\over 4}(1\ldots 1)\cr &\ \cr}$ &
\quad\hbox{none} 
& $\eqalign{&\ \cr & 16\times {1\over 2}\ \hbox{5B} \cr
&\ \cr}$\cr
\noalign{\hrule}
1.3& $V={1\over 4}(1\ldots 1)$ &
\quad\hbox{none} &
$\eqalign{&\ \cr & 8\times {1\over 2}\ \hbox{5B}\cr 
& \hbox{in a 3-plane}\cr
&\hbox{+4 mobile}\ \hbox{5B}\cr &\ \cr}$\cr
\noalign{\hrule}
}
}$$
}}
\endinsert
\vfill\eject
\midinsert{
\vbox{\qquad \qquad \qquad \qquad \qquad
Table B2. $Spin(32)/Z_2$ heterotic orbifolds.
$$\vbox{
\offinterlineskip
\halign{
\strut \vrule \quad $#$\quad &\vrule \hfill \quad #\quad \hfill
&\vrule \quad $#$\quad &\vrule \hfil \quad #\quad \hfill \vrule \cr
\noalign{\hrule}
\hbox{Model} & Shift vector & \hbox{Wilson lines} & Five-branes \cr
\noalign{\hrule}
2.1 & $\eqalign{&\ \cr & V={1\over 4}(1\ldots 1)\cr
&\ \cr}$ &
\eqalign{&\ \cr 
& A_6={1\over 2}(1,1,1,0,0,0,1,0 \cr
&\qquad\qquad 1,1,1,0,0,0,1,0)\cr
& A_7={1\over 2}(1,1,0,1,0,1,0,0 \cr
&\qquad\qquad 1,1,0,1,0,1,0,0)\cr
& A_8={1\over 2}(1,0,1,1,1,0,0,0 \cr
&\qquad\qquad 1,0,1,1,1,0,0,0)\cr
& A_9={1\over 2}(1,1,1,1,1,1,1,1 \cr
&\qquad\qquad 0,0,0,0,0,0,0,0)\cr
&\ \cr} & $\eqalign{ & \hbox{8 mobile} \cr
&\quad \hbox{5B} \cr}$\cr
\noalign{\hrule}
2.2 & $\eqalign{&\ \cr & V={1\over 4}(1\ldots 1)\cr
&\ \cr}$ &
\qquad\qquad\hbox{identical} &
$\eqalign{
& 16\times {1\over 2}\ \hbox{5B} \cr
}$ \cr
\noalign{\hrule}
2.3 & $\eqalign{&\ \cr & V={1\over 4}(1\ldots 1)\cr
&\ \cr}$ &
\qquad\qquad\hbox{identical} &
$\eqalign{&\ \cr
& 8\times {1\over 2}\ \hbox{5B}\cr
&\hbox{in a 3-plane}\cr
& \hbox{+4 mobile}\ \hbox{5B}\cr
&\ \cr}$\cr
\noalign{\hrule}
}
}$$
}}
\endinsert
\vfill\eject
\midinsert{\vbox{\qquad \qquad \qquad \qquad \qquad
Table B3. $Spin(32)/Z_2$ heterotic orbifolds.
$$\vbox{
\offinterlineskip
\halign{
\strut \vrule \quad $#$\quad &\vrule \hfill \quad #\quad \hfill
&\vrule \quad $#$\quad &\vrule \hfil \quad #\quad \hfill \vrule \cr
\noalign{\hrule}
\hbox{Model} & Shift vector & \hbox{Wilson lines} & Five-branes \cr
\noalign{\hrule}
3.1 & $\eqalign{&\ \cr & V={1\over 4}(1\ldots 1)\cr
&\ \cr}$ &
\eqalign{&\ \cr 
& A_6={1\over 2}(1,1,0,0,0,0,0,0\cr
&\qquad\qquad 1,1,0,0,0,0,0,0)\cr
& A_7={1\over 2}(1,0,1,0,0,0,0,0 \cr
&\qquad\qquad 1,0,1,0,0,0,0,0)\cr
& A_8={1\over 2}(1,1,0,0,0,0,0,0\cr
&\qquad\qquad 1,1,0,0,0,0,0,0)\cr
& A_9={1\over 2}(0,0,0,0,0,0,0,0\cr
&\qquad\qquad 0,0,0,0,0,0,0,0)\cr
&\ \cr} & $\eqalign{& \hbox{8 mobile} \cr
&\quad \hbox{5B}\cr}$ \cr
\noalign{\hrule}
3.2 & $\eqalign{&\ \cr & V={1\over 4}(1\ldots 1)\cr
&\ \cr}$ &
\qquad\qquad\hbox{identical} &
$\eqalign{
& 16\times {1\over 2}\ \hbox{5B} \cr
}$ \cr
\noalign{\hrule}
3.3 & $\eqalign{&\ \cr & V={1\over 4}(1\ldots 1)\cr
&\ \cr}$ &
\qquad\qquad\hbox{identical} &
$\eqalign{&\ \cr
& 8\times {1\over 2}\ \hbox{5B}\cr
&\hbox{in a 3-plane}\cr
& \hbox{+4 mobile}\ \hbox{5B}\cr
&\ \cr}$\cr
\noalign{\hrule}
}
}$$
}}
\endinsert
\vfill\eject

\listrefs

\end